\documentclass[a4paper,11pt]{article}
\usepackage{amsmath}
\usepackage{amsfonts}
\usepackage{amssymb}
\usepackage[utf8x]{inputenc}
\usepackage{ae}
\usepackage[T1]{fontenc}
\usepackage[english]{babel}
\usepackage{graphicx}
\usepackage{placeins}
\usepackage{tabularx}
\usepackage{tabulary}
\usepackage{setspace}
\usepackage{tablefootnote}
\usepackage[flushmargin]{footmisc}
\usepackage[comma]{natbib}
\usepackage{array}
\usepackage{float}
\usepackage{caption}
\usepackage{tikz}
\usetikzlibrary{shapes.geometric}
\usepackage[a4paper,left=2.5cm,right=2.5cm,top=3cm,bottom=3cm]{geometry}
\usepackage[colorinlistoftodos]{todonotes}
\usepackage{pdfpages}
\usepackage{longtable}
\usepackage{adjustbox}
\usepackage{booktabs}
\usepackage{multirow}
\usepackage{lscape}
\usepackage[colorlinks=true, allcolors=blue]{hyperref}

\usepackage[flushleft]{threeparttable}
\usepackage{inputenc}



\begin{document} \doublespacing \pagestyle{plain}
	
	\def\ci{\perp\!\!\!\perp}
	\begin{center}
		
		{\LARGE Detecting bid-rigging coalitions in different countries and auction formats}
		
		{\large \vspace{0.8cm}}
		
		{\large David Imhof* and Hannes Wallimann** }\medskip
		
		{\small {*Swiss Competition Commission; University of Fribourg, Dept.\ of Economics; Unidistance, Faculty of Economy, Switzerland \\ **Lucerne University of Applied Sciences and Arts, Competence Center for Mobility; University of Fribourg, Dept.\ of Economics} \bigskip }
	\end{center}
	
	\smallskip

	\noindent \textbf{Abstract:} {\small We propose an original application of screening methods using machine learning to detect collusive groups of firms in procurement auctions. As a methodical innovation, we calculate coalition-based screens by forming coalitions of bidders in tenders to flag bid-rigging cartels. Using Swiss, Japanese and Italian procurement data, we investigate the effectiveness of our method in different countries and auction settings, in our cases first-price sealed-bid and mean-price sealed-bid auctions. We correctly classify 90\% of the collusive and competitive coalitions when applying four machine learning algorithms: lasso, support vector machine, random forest, and super learner ensemble method. Finally, we find that coalition-based screens for the variance and the uniformity of bids are in all the cases the most important predictors according the random forest.
	}
	
	{\small \smallskip }
	{\small \smallskip }
	{\small \smallskip }
	

	{\small \smallskip }
	{\small \smallskip }
	{\small \smallskip }
	

	{\small \smallskip }
	{\small \smallskip }
	{\small \smallskip }
	
	{\small \noindent \textbf{Acknowledgments:}  We have benefited from comments by Christoph Brunner, Martin Huber, Yavuz Karagök, Philippe Sulger, Niklaus Wallimann and Philipp Wegelin.}

	{\small \smallskip }
	{\small \smallskip }
	{\small \smallskip }
	
	{\small \noindent \textbf{Disclaimer:}  All views contained in this paper are solely those of the authors and cannot be attributed to the Swiss Competition Commission, its Secretariat, the University of Fribourg, the Unidistance (Switzerland), or the Lucerne University of Applied Science and Arts.}
	
	{\small \smallskip }
	{\small \smallskip }
	{\small \smallskip }
	
	{\small \noindent  {\scriptsize 
			Addresses for correspondence: David Imhof, Hallwylstrasse 4, 3003 Bern, Switzerland, david.imhof5@gmail.com; Hannes Wallimann, Rösslimatte 48, 6002 Lucerne, Switzerland, hannes.wallimann@hslu.ch.
		}\thispagestyle{empty}\pagebreak  }

	{\small \renewcommand{\thefootnote}{\arabic{footnote}} %
		\setcounter{footnote}{0}  \pagebreak \setcounter{footnote}{0} \pagebreak %
		\setcounter{page}{1} }
	
	\section{Introduction}\label{introduction}
	Bid rigging conspiracies cost governments and taxpayers billions of dollars every year, given that OECD countries spend about 12\% of their GDP on public procurement.\footnote{See https://www.oecd.org/competition/cartels/fightingbidrigginginpublicprocurement.htm (accessed 8 April 2021).} According to the OECD, the elimination of bid rigging could help reduce procurement prices by 20\% or even more. Developing pro-active methods for uncovering bid-rigging conspiracies is therefore of prime importance for competition and procurement agencies all over the world. Pro-active statistical methods to detect bid rigging in public procurement have initially been proposed by, for example, \cite{Harrington2008} and \cite{Porter1993}. The more recent literature discusses the application of a wide range of methods to expose bid-rigging cartels in Brazil \citep[][]{lima2021using}, Canada \citep[][]{clark2018bid}, Japan \citep[][]{Chassang2020}, Sweden \citep[][]{bergman2020interactions} and Switzerland \citep[][]{huber2019machine,imhof2019detecting}.
	
	In this paper, we add to this literature by proposing an original method of detection that focuses on coalitions formed by groups of firms. We apply our method to three different data sets from Japan, Switzerland and Italy for which the incidence of bid rigging is known. In all three countries, we find that on average our method correctly classifies nine coalitions out of ten as collusive or competitive. The results remain robust in different auction formats, such as first-price sealed-bid procurement mechanism in Japan and Switzerland and the mean-price sealed-bid auction in Italy. Our suggested method of detection is thus able to flag collusive groups of firms (collusive coalitions) from different bid-rigging cartels: (i) when all firms in a tender rig the contract, as in Japan and Switzerland \citep[][]{ishii2014bid,huber2019machine,imhof2019detecting}; (ii) when collusive firms face competitive firms, as in Italy and in Switzerland \citep[see][]{conley2016detecting,wallimann2020machine}; and (iii) when a cartel is active mostly in only one region of a market, and the firms rig only a subset of contracts \citep[see][]{imhof2018screening}. 
	
	Our method of detection is based on screens, that is, statistics derived from the distribution of bids in a tender. To derive screens for coalitions, we start by selecting three firms and isolate all the tenders in which those three firms submitted a bid. For each tender, we calculate the screens based exclusively on the three bids of those firms obtaining the tender-based screens for a coalition. We then calculate the descriptive statistics of the tender-based screens, including the mean, median, minimum and maximum for each coalition. These statistics, henceforth called 'coalition-based screens', synthetsize the distributional features of bids for a specific coalition. Since we use data from different bid-rigging cases with complete information, we can identify a coalition as competitive and collusive in order to build the outcome variable. We focus on coalitions of three firms since we aim to detect even small bid-rigging cartels. Focusing on coalitions of two firms would impede the application of most of our screens, and with coalitions of four firms or more it would hinder the detection of the smallest bid-rigging cartels formed by three firms. 
	
	As in recent studies \citep[][]{foremny2018collusion,rabuzin2019prediction,garcia2020bidders,silveira2021won}, we use machine learning to train and test models to flag bid-rigging cartels. For this purpose machine learning is ideal, since it focuses on developing predictive models to determine an outcome. Machine learning does not focus on the causal structural relationship, e.g., between collusion and the distributional pattern of bids. In other words, we remain agnostic about the effects of bid rigging on the distribution of bids when using machine learning techniques. However, we discuss the effects of bid rigging on coalition-based screens by illustrating some common important predictors in all the cases being considered. In our study, we combine the coalition-based screens described above with machine learning to predict whether a coalition of firms colluded in bidding or not. To train predictive models and evaluate their goodness of fit in independent test sets, we apply four widely used machine learning algorithms: the random forest \citep[][]{Breiman2001}, the lasso \citep[][]{frank1993statistical,Tibshirani1996}, the support vector machines \citep[][]{cortes1995support}, and the "super learner" ensemble method, including random forest, neural networks, gradient boosting, and least absolute shrinkage and selection operator (lasso) regression \citep[][]{Laan2008}. 
	
	We first apply our coalition-based approach to the Okinawa bid-rigging cartel from Japan \citep[see also][]{ishii2014bid,huber2020transnational}. The four machine learning algorithms offer correct classification rates from 91.9\% to 94.9\% to classify a coalition as collusive or competitive. In addition, changing the perspective from a tender-based approach to a coalition-based approach increases the correct classification rate of three to six percentage points, corresponding to a decrease of between 27\% and 55\% in the error rate, defined as one minus the correct classification rate. Secondly, we implement our coalition-based approach on Swiss bid-rigging cartels \citep[see also][]{huber2019machine,wallimann2020machine}, and we find correct classification rates from 86.9\% to 90.5\%. The increase in the correct classification rates using a coalition-based approach amounts to four to seven percentage points when comparing the results of the various models applied to complete bid-rigging cartels, which in \cite{wallimann2020machine} amounts approximately to 83\%. Our coalition-based approach therefore reduces the error rate by between 23\% and 44\%, inclusive. Finally, we apply our coalition-based approach to Italian bid-rigging cartels \citep[see also][]{conley2016detecting} and find correct classification rates from 84.8\% to 90.1\% for flagging collusive coalitions. For the three different countries, we find that the medians of the coefficient of variation, the spread and the KS-statistic are the most powerful predictors for flagging collusive coalitions. While the levels of the medians differ strongly between the cases, the effect of bid rigging on the screens goes in the same direction, and its magnitude is to a certain extent similar. Therefore, benchmarks in screening other markets in other countries should rely on the effect of bid rigging. For example, a decrease by a factor of two in the medians of the spread and the coefficient of variation would indicate potential competitive issues requiring further scrutiny.
	
	We complement our analyses in three steps using the Swiss data. First, we add more summary statistics for the tender-based screens. With an enlarged set of coalitions-based screens, we find no significant improvement in the correct prediction rate, indicating that summary statistics based on the mean, median, minimum and maximum are sufficient to deliver a good performance in predicting collusive and competitive coalitions. Second, we discuss why coalition-based screens for the variance and the uniformity of bids perform significantly better than those for the asymmetry of bids. We find that applying only screens for the asymmetry of bids to the Swiss data (omitting the coalition-based screens for the variance and uniformity of bids) produce a poor correct prediction rate. This might be due to the fact that, by forming a coalition (with few firms), the bid of the designated winner and thus the distance between the winning bid and the second lowest bid from the cartel is not systematically considered. Therefore, the asymmetry in the coalition's distribution of bids decreases. Finally, we investigate the number of bidders in coalitions formed with four firms. The result indicates an increase in the correct prediction rates, especially for the collusive coalitions. This might be explained by the increase in predictive power of coalition-based screens for asymmetry. Including more firms in a coalition reduces the likelihood that the first bid in the tender will be omitted. Thus, coalition-based screens appear to be more asymmetric and thus more predictive. 
	
	Our paper relates to other studies using screens for uncovering cartels \citep[see][]{Abrantes2006, Esposito2006, Hueschelrath2011, Jimenez2012, Abrantes2012, huber2019machine, imhof2019detecting}. Calculating screens for subgroups as in our approach is also discussed by \citet{conley2016detecting} and \citet{Chassang2020}. First, \citet{conley2016detecting} calculate subgroups to detect cartels in collusive auctions in Italy. In order to identify collusive bidders, we similarly rely on the bids observed in a tender. However, we do not consider firm-specific covariates, such as common owner, municipality or country, to determine subgroups, as proposed in the study by \citet{conley2016detecting}. The advantage of our method is that we do not rely on firm-specific covariates, which could impede screening activity if firm-specific data are unavailable or if there is not enough time to collect them in secrecy without attracting the attention of potential cartel participants. \citet{Chassang2020} show that winning bids tend to be isolated when bidders collude. They calculate the difference between a bidder’s own bid and the lowest bid submitted in a tender, therefore focusing on subgroups of two bids to calculate the distribution of differences. However, we do not focus solely on subgroups consisting of only the lowest bid in a tender and one of its opposing bidders.
	
	More broadly, our study can be linked to papers on detecting bid-rigging cartels not relying on screens. One seminal paper by \citet{Bajari2003} proposes two econometric tests for classifying pairs of firms as collusive. Subsequent papers apply and refine the econometric tests suggested by Bajari and Ye (2003) \citep[see][]{Jakobsson2007, Aryal2013, Chotibhongs2012a, Chotibhongs2012b, Imhof2017b, bergman2020interactions}. \citet{Imhof2017b}, however, questions the performance of the econometric tests proposed by \citet{Bajari2003} for detecting the Ticino cartel because econometric tests produce too many false negatives by failing to classify pairs of firms as collusive, whereas the screens perform well in detecting the Ticino cartel. Our research is also associated with papers analyzing the effect of bid rigging \citep[][]{Pesendorfer2000, Ishii2009, clark2018bid} and to papers investigating the change in bidding patterns when bid rigging occurs \citep[][]{Porter1993, Porter1999}.
	
	The remainder of the paper is organized as follows. Section \ref{detecmeth} outlines our method of detection. In Section \ref{empanaly}, we apply our detection method to public procurement datasets from Italy, Japan and Switzerland. We also discuss the observed variance screens and the Kolmogorov-Smirnov statistic, which are important in flagging bid-rigging cartels. Section \ref{companaly} performs complementary analyses. In Section \ref{policyrecom}, we discuss the advantages and policy implications of our approach. Section \ref{Conclusion} concludes the paper. 
	
	\section{Detection method}\label{detecmeth}
	
	In our study, we focus on supervised machine learning that entails a set of predictors $(X)$, also features or covariates, to predict an outcome $(Y)$. The outcome of our classification setting is given a value of 1 for a collusive coalition, which only includes cartel participants, and a value of 0 for a competitive coalition, which is formed only by competing firms. Machine learning requires the data to be randomly split into independent training and test datasets. In our applications, the training and test sets consist of 75\% and 25\% of the observations respectively. We develop predictive models using all observations in our training set, where both features and outcomes are observed. The goal is to predict for each observation the outcomes in the test data on the basis of their covariates. This is closely related to discrete choice analysis in econometrics, where statistical models specify a probability that an outcome takes a particular value conditional on the features \citep[][]{athey2019machine}. However, machine learning aims to achieve goodness of fit in the independent test set by minimizing deviations between the predicted and the actual outcomes \citep[][]{athey2019}. 
	We assess the predictive performance of machine learning algorithms by comparing the prediction of the algorithm with the actual outcome in the test set. The number of correct predictions divided by the total number of observations in the test set defines the ‘accuracy’ (also the correct classification rate) achieved by the algorithms. For every application, we create a dataset in which the binary outcome is balanced, i.e., with 50\% collusive and 50\% competitive coalitions. Balancing the dataset each time before splitting the sample enables the applied algorithms to build models predicting both coalition classes, collusive and competitive, equally well. After randomly balancing the dataset, we repeat the sample splitting into training and test data a hundred times. The correct classification rates of our applications are the average predictive performances of the hundred repetitions. 
	In our study, we train machine learning algorithms with coalition-based screens $(X)$ to flag collusive coalitions $(Y)$ in the three countries of Italy, Switzerland and Japan. In the following, we first discuss the machine learning algorithms used for training and testing our predictive models. We then describe the coalition-based approach and the screens entering in the algorithms as features. 
	
	\subsection{Machine learning algorithms}
	
	The first machine learning algorithm we implement is the least absolute shrinkage and selection operator (lasso) regression, introduced by Frank and \cite{frank1993statistical} and \cite{Tibshirani1996}. In our case a lasso regression is a type of logit regression using shrinkage. It includes a penalty term, restricting the sum of absolute coefficients on the regressors. Coefficients with a low predictive power shrink, depending on the penalty term, towards or exactly to zero. As some coefficients become zero and the algorithm discards these variables from the model, the lasso regression can result in sparse models with only the most powerful predictive variables. Based on the mean squared error of prediction, we apply a 15-fold cross-validation to select the penalty term. In our applications of the lasso regression, we use the \textit{hdm} package by \cite{Chern2016} in the statistical software R. 
	
	Second, we use the random forest \citep[see][]{Breiman2001}, an algorithm predicting the outcome by a majority rule across multiple individual decision trees. This machine learning method therefore draws random subsamples from the original training set and estimates the predictive model, in our case a decision tree, in each of the subsamples. A decision tree splits the feature space into a number of non-overlapping regions. Each split aims to maximize the homogeneity of the outcome according to a goodness of fit criterion. In the case of binary variables, the Gini coefficient is a popular criterion that measures the average gain in homogeneity of the outcome values. The splitting continues until the tree reaches a specific stopping rule, e.g., the minimum number of observations in a terminal node. The tree-based predictions of the outcome are based on whether collusive coalitions are present or absent in the region that contains the values of features for which the outcome is to be predicted. Using tree-based methods, there exist a bias-variance trade-off in the out-of-sample prediction. Through more splits, on the one hand, we reduce the bias and increase the flexibility of the model specification. On the other hand, more splits increase the variance in the test data due to regions being smaller. By repeatedly drawing many subsamples from the training set and estimating the decision tree, the random forest mitigates the issue of excessive variance in the test set. To reduce the correlation of tree structures across the subsamples and the prediction variation, the random forest considers at each splitting step of each decision tree only a random subsample of features. In our applications, the subsample of features at each split amounts to the square root of potential predictors. To implement the random forest in the statistical software R, we use the \textit{randomForest} package of \cite{Liaw2018randomForest}, with growing a thousand trees. Implementing the random forest, we also present the most important variables according to the Gini Index as a measure of the best split selection, which measures the impurity of a given element with respect to the remaining classes. 
	
	Third, we implement support vector machines \citep[][]{cortes1995support}. Support vector classifiers are based on the idea of finding a hyperplane that best segregates the training data into two categories. We can think of a hyperplane as a line separating the observed points in a two-dimensional space into two classes. We then map the observations of the test data into the space and predict them to belong to one class based on the side of the hyperplane on which they fall. In the training data, we want our data points to be as far away as possible from the hyperplane, as for these data points confidence in their producing a correct classification will be high. The distance from the nearest data point in either of the two separated classes and the hyperplane is known as the margin. Giving a greater chance of new data being correctly classified, the algorithm choses a hyperplane with the goal of achieving the greatest possible margin. However, the idea of the hyperplane as a line is a simplification, as a linear hyperplane might perform poorly when the data points are not separable with a line. Support vector machines offer an extension of the support vector classifier by enlarging the feature space using kernels and mapping the inputs into high-dimensional feature spaces. In our application of support vector machines, we use the \textit{e1071} package \cite{meyer2015support}.
	
	Fourth, we apply the \textit{SuperLearner} package by \cite{Laan2008}, which is an ensemble method. In our case, the super learner is a weighted average of four machine learning algorithms: gradient boosting, random forest, lasso and neural networks, using the \textit{xgboost}, \textit{cforest}, \textit{glmnet} and \textit{nnet} packages respectively. Gradient boosting resembles the random forest described above, as it grows a set of decision trees. However, unlike the later algorithm building each tree independently, gradient boosting is an additive model working in a forward stage-wise manner and therefore building only one tree at a time. While the random forest averages over all decision trees at the end, gradient boosting combines the results along the way. That is, building individual decision trees sequentially learning from mistakes made by previous ones. Neural networks aim at fitting a system of nonlinear functions modelling the influence of the features on the outcome in a flexible way. The algorithm uses a network of non-linear intermediate functions, so-called hidden nodes, to model the association between the predictors and the outcomes.

	\subsection{Coalitions and predictors}
	
	Procurement markets are seldomly characterized by a complete cartel involving all firms bidding for a tender. In such cases, a suspicious group of bidders (a coalition) must be isolated by further statistical tests, as suggested, for example, by \cite{imhof2018screening}. As a methodical innovation, in our paper we develop a coalition-based approach for flagging cartel participants. Our approach overcomes the isolation step and directly identifies collusive coalitions. By 'coalition' we mean a subgroup of three firms bidding together in at least three tenders. Coalition-based screens consisting of three bidders are able to detect bid-rigging cartels including more than three firms. However, coalition-based screens consisting of four bidders would not be able to flag bid-rigging cartels including with only three bidders. We therefore focus on three firms since our main aim is to create a method of detection that is capable of flagging small bid-rigging cartels as well as large ones.\footnote{Using coalitions of only two firms does not allow us to calculate all possible screens, leaving us with a reduced set of predictors.} 
	
\begin{figure}[!htp]
		\centering \caption{\label{Coalition123} The selection of coalition 123}
		\includegraphics[scale=0.45]{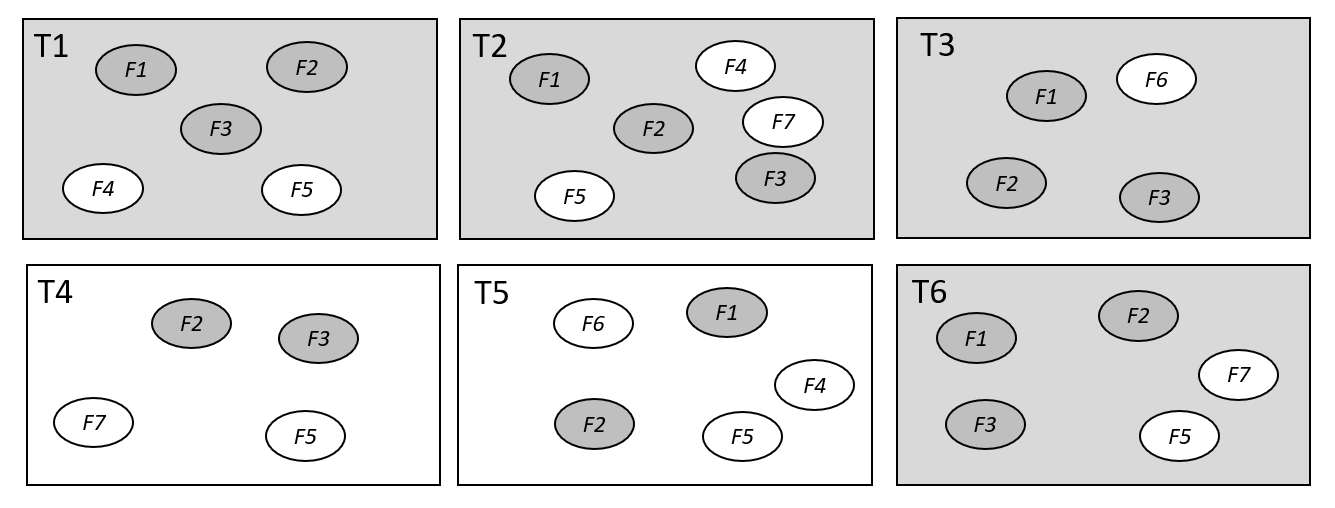}
	\end{figure}

	To prepare the predictors for an observation (a coalition), we extract all the tenders in which each of the three firms submitted a bid and discard all bids submitted by firms that are not part of the coalition. Figure \ref{Coalition123} illustrates the procedure. The boxes represent tenders, the circles firms. In Figure \ref{Coalition123}, we have a sample with six tenders T1 to T6 and seven firms F1 to F7 applying for projects. To form the first coalition, we pick firms F1, F2 and F3, henceforth called coalition 123. Each of these firms submits a bid in each of the tenders T1, T2, T3 and T6 (in grey). To form coalition 123, we extract this subgroup of tenders and discard all bids submitted by firms that are not part of the coalition. For example, in tender T1 we drop firms F4 and F5 (white circles).

	In a next step, we calculate screens for the distribution of the three extracted bids in each tender. Screens are descriptive statistics describing the discrete distribution of bids in a tender \citep[][]{Abrantes2006,Abrantes2012,Harrington2008,Jimenez2012,imhof2019detecting}. Since screens summarize the behavior of the bidders in one tender, they refer to the category of behavioral screens as discussed by \cite{Harrington2008}. We make the simple hypothesis that bid rigging modifies the distribution of bids. There are two reasons for this: (i) the members of a bid rigging cartel know the bids of their competitors, and (ii) they coordinate the bids. Therefore, we can capture such distributional changes with the screens. In fact, this hypothesis is common to detection methods such as the econometric tests suggested by \cite{Bajari2003}. 
	
	Following \cite{huber2020transnational}, \cite{huber2019machine} and \cite{wallimann2020machine}, we implement nine screens to uncover bid-rigging cartels. The screens can be assigned to three categories. The first category contains variance screens such as the coefficient of variation \citep[see e.g.][]{Abrantes2006,Abrantes2012,imhof2019detecting,Jimenez2012} and the spread \citep[see e.g.][]{wallimann2020machine}. These screens capture the possible reduction in the support of the distribution of bids or the convergence of bids when a cartel coordinates bids, and bidders exchange their bids before submitting them in the tendering process \citep[][]{imhof2019detecting}. The second category contains the percentage difference, the absolute difference, the skewness, the relative distance, the alternative distance and the normalized distance \citep[see for these screens][]{huber2019machine}. Screens of this category measure whether the bids exhibit an asymmetrical distribution. Cartel participants can simultaneously affect both differences between losing bids and differences between the first and second lowest cartel bids. Empirical observations \citep[see e.g.][]{Chassang2020} have shown that the differences between the first and second lowest cartel bids increase, whereas the differences between losing bids decrease. This increases the asymmetry in the distribution of the bids and is explained by the necessity to ensure the contract is awarded to the winner designated by the cartel. The third category of predictors is based on the Kolmogorov-Smirnov statistic (hereafter the KS statistic), which is calculated to test whether the discrete distribution of bids follows a uniform distribution \citep[see][]{wallimann2020machine}. The KS statistic thus investigates how dissimilar the distribution of the bids is with a uniform probability distribution due to bid rigging.
	
	By looking again at coalition 123, we illustrate the calculation of the screens with the coefficient of variation. For each tender in the extracted subgroup of tenders T1, T2, T3 and T6, we first calculate the coefficient of variation, that is, the standard deviation divided by the arithmetic mean of the three bids of firms F1, F2 and F3. We thus obtain four coefficients of variation, in other words four tender-based screens. Thereafter, we calculate the mean, median, minimum and maximum using these tender-based screens in order to obtain summary statistics for each coalition, the so-called coalition-based screens.  Calculating the coalition-based screens for each screen presented above, we end up with 36 coalition-based screens for a coalition (observation) in the data. We then use these coalition-based screens as features $(X)$ in our predictive models to determine the outcome $(Y)$. 
	
	We repeat the building of coalitions and the calculation of the coalition-based screens for all possible coalitions of three firms if the three firms at least participate together in three tenders.\footnote{We have chosen three projects, as this is the minimum for calculating summary statistics and allows us to achieve the most observations possible.}
	
	\section{Empirical analyses in different countries}\label{empanaly}
	
	We apply our original coalition-based approach to uncover collusive cartels in three countries: Japan, Switzerland and Italy. These cases are discussed and screened in earlier studies \citep[see][]{conley2016detecting,wallimann2020machine,huber2019machine,ishii2014bid}. While procurement agencies in Japan and Switzerland used first-price sealed-bid auctions, agencies in Italy used mean-price sealed-bid auctions. Therefore, we are able to train and evaluate our algorithms in different countries as well as different auction settings. For every application, we also present the ten most important variables ranked by the Gini Index according to the random forest. Since the most important coalition-based screens for prediction remain stable across countries and auction settings, we briefly discuss them for further screening applications. 
	
	\subsection{Okinawa cartel}
	
	For our first application we use an empirical dataset from Japan originally introduced by \cite{ishii2014bid} and recently analyzed by \cite{huber2020transnational}. The dataset contains construction contracts in Okinawa from April 2003 to March 2007. As the Okinawa Prefecture consists of 47 islands, the market is difficult to enter for firms outside this region, as there is a natural geographical barrier to their entry into the construction market. To procure a contract, local agencies used a first-price sealed-bid auction and specified a reserve price and a lowest acceptable price. The lowest bid submitted between the lowest acceptable price and the reserve price won the contract.
	
	During the whole period, the agency invited a set of qualified firms to submit a bid depending on the size of the tendered contract. In addition, agencies disclosed the identity of the invited firms prior to each tender procedure, a practice that ended in January 2006. The natural geographical barriers, the restricted number of competitors and the disclosure of their identity notably simplified the emergence of bid rigging. Hence, the cartel participants communicated with each other prior to each tender and met to negotiate and agree on the firm that would win the contract, as well as on the winning price. Thereafter, the other bidders not chosen to win the contract calculated a cover bid that was sufficiently higher than the winning price.
	
	In June 2005, the Japanese Fair Trade Commission (hereafter JFTC) launched an investigation into bid-rigging conspiracies against a large number of firms involved in these tenders. To limit the risk of bid-rigging in the future, in January 2006 the Okinawa prefecture adapted its procurement system by inviting more firms and not revealing the identities of firms prior to the tendering procedure. At the same time, Japan’s competition law was revised. Changes included increasing fines for conspiracies and introducing a leniency program granting complete or partial exemption from financial penalties if a firm collaborates with the JFTC. 
	
	\begin{table}[ht]
		\caption{The correct classification rates for the Okinawa cartel}\label{CCRoki}
		\begin{center}
			\begin{tabular}{lccc}
				\hline
				\multirow{2}{*}{Classifier} & \multicolumn{3}{c}{Prediction Results}               \\ \cline{2-4} 
				& CCR (\%) & CCR collusion (\%) & CCR competition (\%) \\ \hline
				Lasso                       & 92.3    & 93.5              & 91.2                \\
				Random   forest             & 94.9    & 96.9              & 92.8                \\
				Super learner               & 93.9    & 94.7              & 93.1                \\
				Support   vector machines   & 91.9    & 93.7              & 90.0                \\ \hline
			\end{tabular}
		\end{center}
		\par
		\textit{Note: 'CCR' denotes the correct classification rate, 'CCR collusion' the correct classification rate of the collusive coalitions, and 'CCR competition' the correct classification rate of competitive coalitions.}
	\end{table}
	
	To create the Japanese collusive coalitions, we consider contracts of type A+ in the pre-inspection period \citep[see][for more details]{huber2020transnational}. For the competitive coalition, we use only contracts of type A+ in the post-amendment period, in which the JFTC sanctioned the cartel participants involved in light of Japanese competition law being revised and procurement rules in Okinawa reinforced. After recreating the coalitions, our final dataset contains 207 collusive and 1,793 competitive coalitions. The average number of projects per coalition amounts to 3.4 for both the pre- and post-amendment periods.
	
	As stated in Table \ref{CCRoki}, we use the four algorithms presented above to achieve decent correct prediction rates from 91.9\% to 94.9\%. Therefore, the deviations between the predicted and actual outcomes are low. Our coalition-based approach outperforms the application of \cite{huber2020transnational} by about two to six percentage points depending on the algorithm.\footnote{We do not compare our analysis with model 1 in \cite{huber2020transnational} but with model 2, which uses only screens, as in our approach.}  This performance improvement may seem weak, but it is in fact notable if we consider the error rate (also the misclassification error), defined as one minus the correct prediction rate. In \cite{huber2020transnational}, the error rate amounts to approximatively eleven percentage points for models using screens exclusively. An improvement of two to six percentage points implies a reduction of the error rate of between 18\% and 55\% inclusive. Such a reduction in the error rate is valuable in light of the heavy legal consequences of a firm being flagged as potential cartel participant and an investigation being opened against it. Furthermore, investigation has procedural consequences in being costly for both the authority, i.e. the taxpayer, and the firms. 
	
	By comparing the accuracy of the machine learning algorithms, we see from Table \ref{CCRoki} that the random forest achieves the highest correct classification rate. For all algorithms, differences in the predictive performance between the collusive and competitive coalitions remain minor, despite slightly better prediction rates for the collusive coalitions. Yet, the imbalance is the smallest for the super learner. 
	
	\begin{figure}[!htp]
		\centering \caption{\label{VarImp_Oki} Variable importance plot for the Okinawa cartel. We compute the variable importance using the mean decrease in the Gini index and express it relative to the maximum.}
		\includegraphics[scale=0.45]{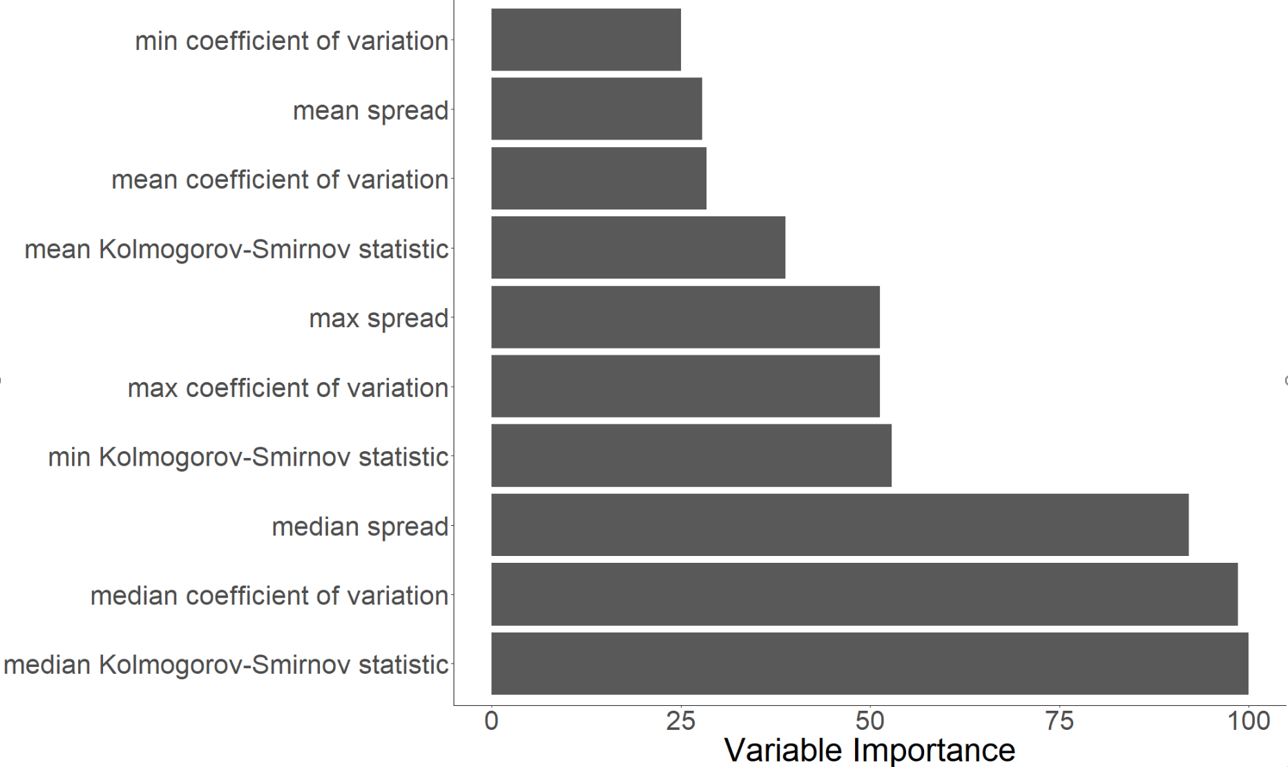}
	\end{figure}
	
	Figure \ref{VarImp_Oki} depicts the relative importance of the predictors according to the random forest. We notice that the median of the coefficient of variation, the spread and the KS statistic are the most important coalition-based screens for predicting bid rigging. Screens for asymmetry, however, appear to be unimportant in predicting bid rigging when also using screens for the variance or uniformity of bids to fit machine learning models. 
	
	\subsection{Swiss cartels}
	
	Our second application considers the dataset from three bid-rigging cartels in Ticino, See-Gaster and Graubünden, discussed by \cite{wallimann2020machine}. The Swiss Competition Commission (hereafter COMCO) convicted cartel participants in all three cases. COMCO only sanctioned cartel participants in two cases since the bid-rigging cartel in Ticino ceased its illegal activity before the revised competition law in Switzerland entered into force, including the possibility of sanctioning firms. The latter cartel was active in the period from January 1999 to March 2005 and included all firms in the road construction market in Switzerland’s southernmost canton \citep[see also][]{imhof2019detecting}. The firms rigged public and private contracts before stopping their anticompetitive activity. After the cartel came to an end, prices fell by roughly 30\% \citep[][]{imhof2019detecting}.
	
	From 2004 to 2010, eight firms in the See-Gaster region (cantons of St. Gallen and Schwyz) participated in a bid-rigging conspiracy. The cartel participants met at least once a month to discuss future tenders for road construction, asphalting and civil engineering. The cartel members designated the winning firm, which then negotiated the price itself, and the cover bids with the cartel participants in separate meetings.
	
	The third cartel, which was active from 2004 to 2010, included most of the road construction firms in the canton of Graubünden, a canton characterized by valleys and mountains. The cartel was divided into two groups of cartel participants operating in the north and the south respectively. As in the two latter investigations, the cartel participants discussed local and cantonal contracts for asphalting and construction tendered by the canton and the cities. COMCO estimated that the activities of the cartel pushed up prices by at least 10\%.
	
	In Switzerland, in awarding contracts, procurement agencies also take other criteria into account and not just price, such as quality considerations and environmental aspects. Price, however, remains the most important criterion. Therefore, the procurement process in Switzerland is characterized by a first-price sealed-bid auction \citep[for further explanations, see][]{wallimann2020machine}. 
	
	\begin{table}[ht]
		\caption{The correct classification rates for the Swiss cartels}\label{CCRSwisscart}
		\begin{center}
			\begin{tabular}{lccc}
				\hline
				\multirow{2}{*}{Classifier} & \multicolumn{3}{c}{Prediction Results}               \\ \cline{2-4} 
				& CCR (\%) & CCR collusion (\%) & CCR competition (\%) \\ \hline
				Lasso                       & 86.9    & 88.5              & 85.4                \\
				Random   forest             & 89.7    & 88.3              & 91.1                \\
				Super learner               & 90.5    & 90.0              & 91.1                \\
				Support   vector machines   & 87.2    & 88.1              & 86.3                \\ \hline
			\end{tabular}
		\end{center}
		\par
		\textit{Note: 'CCR' denotes the correct classification rate, 'CCR collusion' the correct classification rate of the collusive coalitions, and 'CCR competition' the correct classification rate of competitive coalitions.}
	\end{table}
	
	\begin{figure}[!htp]
		\centering \caption{\label{VarImp_Swiss} Variable importance plot for the Swiss cartels. We compute the variable importance using the mean decrease in the Gini index and express it relative to the maximum.}
		\includegraphics[scale=0.45]{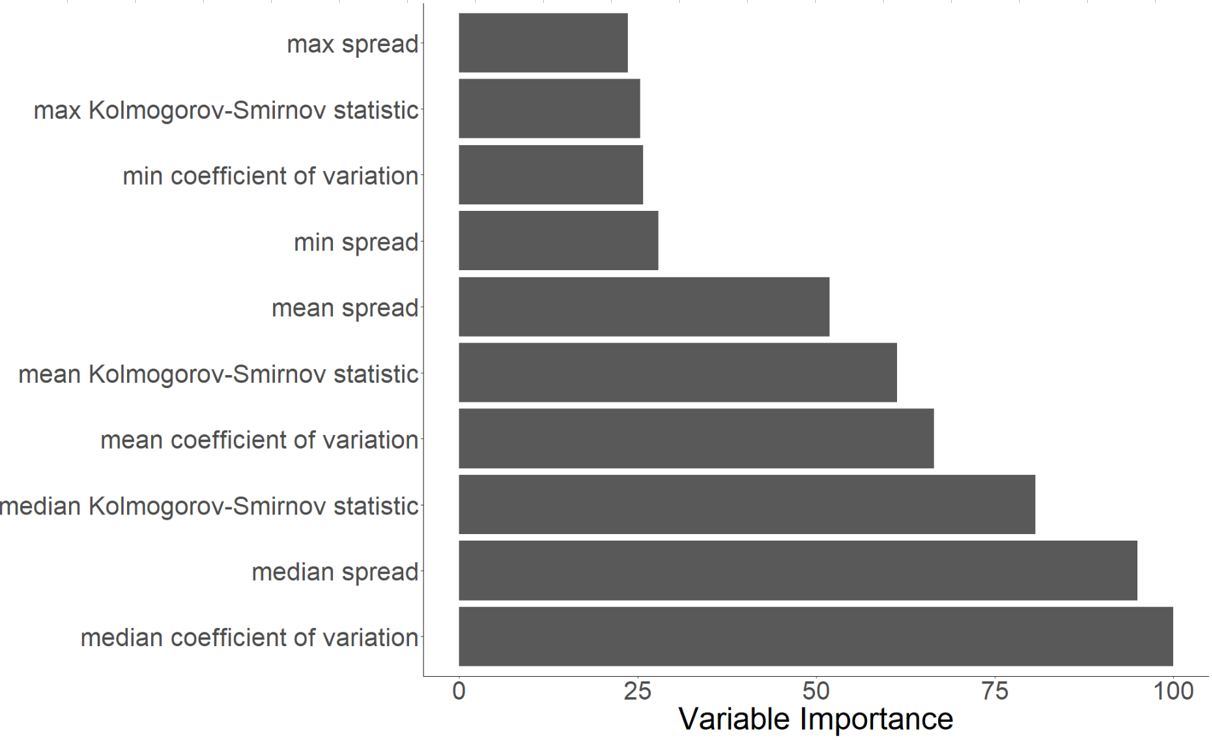}
	\end{figure}
	
	In this study, we pool the data of all three cartels. We use cartel participants to construct collusive coalitions. Competitive coalitions are created with former cartel participants to investigate the changes between the collusive and competitive coalitions. At the end of the formation of all coalitions, we end up with 646 competitive and 896 collusive coalitions. In our data, the average number of projects per coalition amounts to 21.4 and 44.9 for the competitive and collusive coalitions respectively.
	
	As shown in Table \ref{CCRSwisscart}, the correct prediction rates amount to 86.9\%, 89.7\%, 90.5\% and 87.2\% for the lasso, random forest, super learner and support vector machines respectively. The super learner reaches the lowest misclassification error in predicting collusive and competitive coalitions in the Swiss data. We improve the predictive performances of \cite{wallimann2020machine} by three to seven percentage points if we consider only the complete bid-rigging cartels (with no competition of firms that are not part of the cartel) in the various models applied. Such increases in the correct prediction rate implies a decrease of between 23\% and 44\% inclusive in the error rate.
	
	Like the Okinawa application, we observe a convergence of the algorithms but a slightly better performance for the random forest and super learner. We also notice that the random forest and the super learner are slightly better at predicting competitive coalitions, implying that they produce fewer false positives (one minus the correct prediction rate for competitive tenders) than false negatives (wrongly flagging a collusive coalition as competitive). The reverse applies to the lasso and the support vector machines, which predict better collusive coalitions but with a lower overall correct classification rate. In fact, all four machine learners exhibit a similar correct classification rate for the collusive coalition, i.e. the same false negative results. Increasing false positive results for the lasso and the support vector machines explains the difference in the overall correct classification rates.
	
	Figure \ref{VarImp_Swiss} reports the most important coalition-based screens according to the random forest. We again observe that medians for the KS statistic and the spread and coefficient of variation are the three most important coalition-based screens in classifying collusive coalitions. Again, in the ten most predictive coalition-based features we do not find any screens for the asymmetry of bids as for the Okinawa application.

	\subsection{Italian cartels}
	
	Our third application involves contracts for roadworks tendered in the Turin municipality of Italy between 2000 and 2003, first introduced by \cite{conley2016detecting}. They use two datasets: a validation dataset and a main dataset. In our application, we use the validation dataset because there are no court decisions in the main dataset and we thus have no prior knowledge on the existence of a cartel in this dataset.
	
	The procurement agency in Turin used an average bid auction for tendering the roadwork contracts. First, it defined a reserve price for a contract and publicly announced it. Based on the reserve price, interested firms submitted a bid, which was a discount based on the reserve price. Having collected all the bids, the agency first ranked them and discarded the ten percent lowest and highest bids to calculate a trimmed mean. The agency then calculated a second mean for all bids (including discarded ones) higher than the trimmed mean in the first step. The firm with the highest bid lower than the mean of the second step won the contract \citep[see][for details]{conley2016detecting}. 
	
	In 2008, the Court of Justice in Turin identified eight cartels involving 95 firms as potential cartel participants and sentenced 27 firms for bid-rigging conspiracies. The firms mostly formed cartels with nearby companies. Overall, the coordination of bids paid off because the suspected cartel participants won 80\% of the tendered contracts, though they accounted for only 10\% of all the bidders.

	\begin{table}[ht]
		\caption{The correct classification rates for the Italian cartels}\label{CCRItalcart}
		\begin{center}
			\begin{tabular}{lccc}
				\hline
				\multirow{2}{*}{Classifier} & \multicolumn{3}{c}{Prediction Results}               \\ \cline{2-4} 
				& CCR (\%) & CCR collusion (\%) & CCR competition (\%) \\ \hline
				Lasso                       & 84.8    & 83.9              & 85.8                \\
				Random   Forest             & 89.1    & 87.6              & 90.6                \\
				Super learner               & 90.1    & 89.9              & 90.3                \\
				Support   Vector Machines   & 85.2    & 83.2              & 87.3                \\ \hline
			\end{tabular}
		\end{center}
		\par
		\textit{Note: 'CCR' denotes the correct classification rate, 'CCR collusion' the correct classification rate of the collusive coalitions, and 'CCR competition' the correct classification rate of competitive coalitions.}
	\end{table}
	
	By recreating the coalitions in the Italian data, we take 75 of the most frequent competitive bidders and obtain 21,340 competitive coalitions with an average of 20.7 contracts. We calculate collusive coalitions within each of the eight cartels.  We end up with 1,474 collusive coalitions with an average of 47.4 contracts.
	
	Our coalition-based models reach correct classification rates from 84.8\% to 90.1\% in detecting the Italian cartels and therefore perform well in a different kind of auction procedure (see Table \ref{CCRItalcart}). Again, we find the super learner and the random forest to be the best performing algorithms compared to the lasso and the support vector machines. We notice that the lasso, the support vector machines and the random forest perform better in predicting competitive coalitions and thus produce fewer false negatives than false positives. 
	
\begin{figure}[!htp]
		\centering \caption{\label{VarImp_It} Variable importance plot for the Italian cartels. We compute the variable importance using the mean decrease in the Gini index and express it relative to the maximum.}
		\includegraphics[scale=.45]{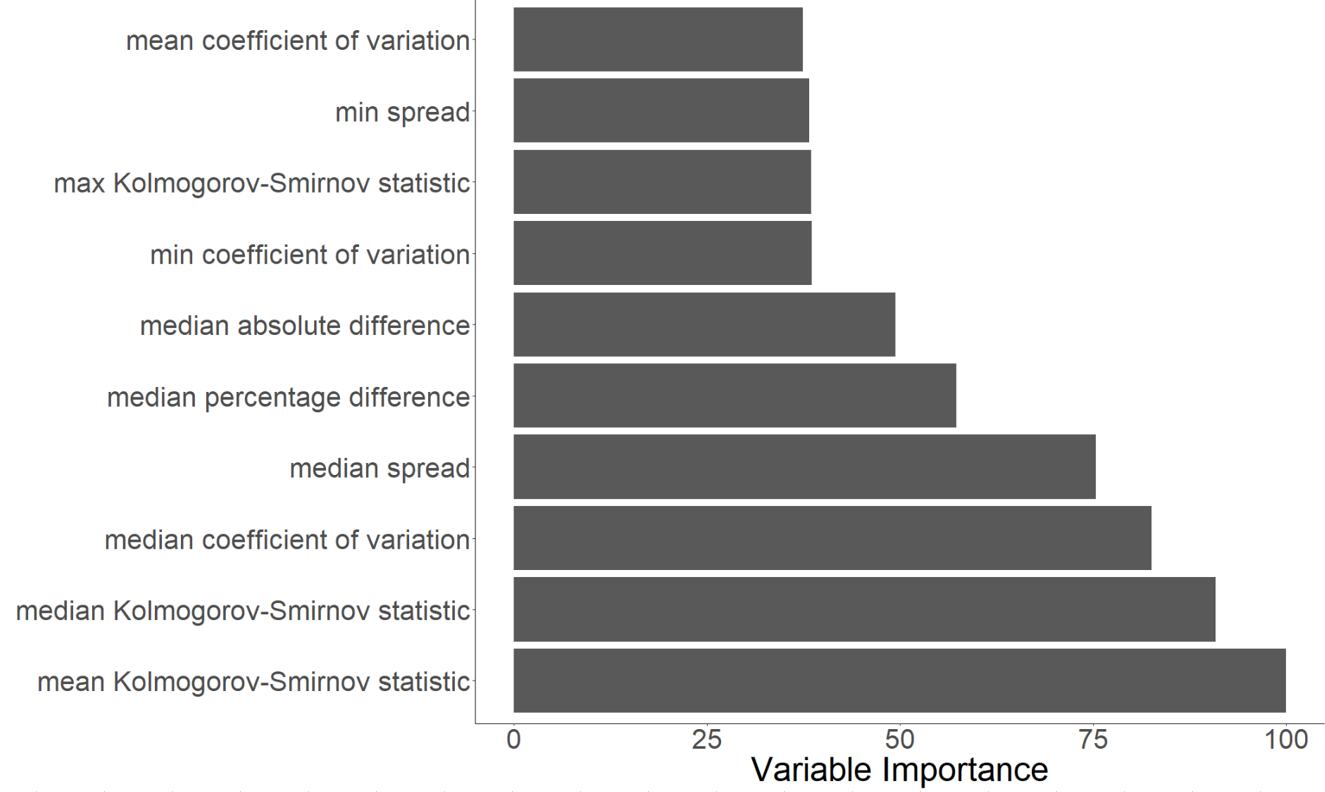}
	\end{figure}

	Figure \ref{VarImp_It} presents the ten most predictive coalition-based screens in predicting Italian collusive coalitions. We again find similar important predictors as in the latter applications: the median for the KS statistic and the coefficient of variation and spread are the most predictive coalition-based screens with the mean of the KS statistic. Unlike the previous cases, we notice two screens for the asymmetry of bids, with the median of the percentage difference and of the absolute difference also playing a role in the top-ten predictors. Nevertheless, screens for the variance and uniformity of bids clearly dominate the best predictors.

	\subsection{The most predictive coalition-based screens}
	
	Applying our approach to three different countries, we find the same screens to be the most important predictors $(X)$ for flagging collusive coalitions $(Y)$: we mainly find coalition-based screens for the variance, i.e., medians of the coefficient of variation and the spread. Table \ref{Coalitionsmeans} reports the mean values of these coalition-based screens. We find that the medians of the spread and of the coefficient of variation are on average considerably higher for competitive coalitions. If the level of the variance of bids differs across countries, the effect of bid rigging is similar in magnitude. Bid rigging affects the variance of the bids by decreasing them by two for Swiss bid-rigging cartels and by three for the Italian and Japanese bid-rigging cartels. Bid rigging also decreases the spread by a factor of two for Switzerland, by a factor of three for Japan and by a factor of between three and four for Italy.

	\begin{table}[ht]
		\caption{Mean and standard deviation of the coalitions' medians}\label{Coalitionsmeans}
		\begin{center}
			\begin{tabular}{lcccccccc}
				\hline
				& \multicolumn{2}{c}{Okinawa} & \multicolumn{1}{l}{} & \multicolumn{2}{c}{Italy} & \multicolumn{1}{l}{} & \multicolumn{2}{c}{Switzerland} \\ \cline{2-3} \cline{5-6} \cline{8-9} 
				& Coll.        & Comp.        &                      & Coll.       & Comp.       &                      & Coll.           & Comp.         \\ \hline
				\multirow{2}{*}{Coefficient of variation} & 1.06         & 3.19         &                      & 10.13       & 30.73       &                      & 3.38            & 6.80          \\
				& (2.48)       & (2.86)       &                      & (16.17)     & (20.68)     &                      & (1.58)          & (3.01)        \\
				\multirow{2}{*}{Spread }                                   & 0.02         & 0.06         &                      & 0.32        & 1.16        &                      & 0.07            & 0.14          \\
				& (0.05)       & (0.06)       &                      & (0.70)      & (1.25)      &                      & (0.02)          & (0.07)        \\
				\multirow{2}{*}{KS statistic}              & 286.53       & 143.60       &                      & 35.57       & 7.13        &                      & 34.22           & 17.81         \\
				& (192.43)     & (559.20)     &                      & (40.85)     & (9.72)      &                      & (15.17)         & (7.03)        \\ \hline
			\end{tabular}
		\end{center}
		\par
		\textit{Note: 'Coll.' denotes collusive coalitions, 'Comp.' competitive coalitions. The figures in brackets are the standard deviations.} 
	\end{table}
	
	Alongside with screens for variance, we find that the median of the KS statistic, calculated to test if a discrete bid distribution follows a uniform probability distribution law, is also a powerful coalition-based screen. Table \ref{Coalitionsmeans} indicates that bid rigging notably increases the KS statistic in all cases. In other words, the results suggest that bid rigging and the related necessary bid coordination transform the distribution of bids in a much less uniform distribution. Again, the level of the KS statistic differs across countries, but the effect of bid rigging follows the same direction in all cases. Bid rigging on average doubles the KS statistic for coalitions in Japan and Switzerland compared to their competitive counterparts, whereas for the former in Italy, this screen increases by a factor of five. 
	
	\section{Complementary analyses}\label{companaly}
	
	In this section, we outline the complementary analyses we perform using the Swiss data. First, we enlarge our set of predictors. Second, we investigate why coalition-based screens for the variance and uniformity of bids perform better that those for the asymmetry of bids. Finally, we form coalitions with four firms. 
	
	\subsection{Using additional coalition-based screens}
	
	In the previous section, we calculate coalition-based screens $(X)$ by taking into account the summary statistics mean, median, minimum and maximum of the tender-based screens for each coalition. In this section, we investigate the robustness of these summary statistics, chosen using the Swiss data. Therefore, we calculate the 5th, 10th, 25th, 75th, 90th and 95th percentiles from the tender-based screens for each coalition in the Swiss data. We add them to the coalition-based screens we use in our original application, i.e., mean, median, minimum and maximum. Thus, we fit our models with ninety coalition-based screens in our first complementary analysis.
	
	\begin{table}[ht]
		\caption{Changes in accuracy when adding a new set of predictors in the application of the Swiss cartels}\label{CCRnewpredictors}
		\begin{center}
			\begin{tabular}{lccc}
				\hline
				\multirow{2}{*}{Classifier} & \multicolumn{3}{c}{Changes in percentage points}               \\ \cline{2-4} 
				& CCR  & CCR collusion  & CCR competition \\ \hline
				Lasso                       & 1.8     & 0.7               & 2.9                \\
				Random   forest             & -0.1    & -0.8              & 0.6                \\
				Super learner               & 0.6     & -0.1              & 1.3                \\
				Support   vector machines   & 0.9     & 0.1               & 1.7                \\ \hline
			\end{tabular}
		\end{center}
		\par
		\textit{Note: 'CCR' denotes the correct classification rate, 'CCR collusion' the correct classification rate of the collusive coalitions, and 'CCR competition' the correct classification rate of competitive coalitions.}
	\end{table}
	
	Table \ref{CCRnewpredictors} shows the increase in percentage points of the correct classification rates when performing this analysis compared to the results obtained using the Swiss data in Section \ref{empanaly}. We observe that the overall improvement in accuracy is quite low, amounting to from -0.1 to 1.8 percentage points depending on the algorithm. The increase is the highest for the lasso, but slightly negative for the random forest. We notice that the predictive performance increases more for the competitive coalitions (from 0.6 to 2.9 percentage points), while remaining stable for the collusive coalitions (from -0.8 to 0.7 percentage points). As the overall change in goodness of fit for the four algorithms is quite low, we assume that the gain of additional coalition-based screens is negligible. 
	
	\begin{figure}[!htp]
		\centering \caption{\label{VarImp_MorePred} Variable importance plot for the Swiss cartels with more predictors. We compute the variable importance using the mean decrease in the Gini index and express it relative to the maximum.}
		\includegraphics[scale=0.45]{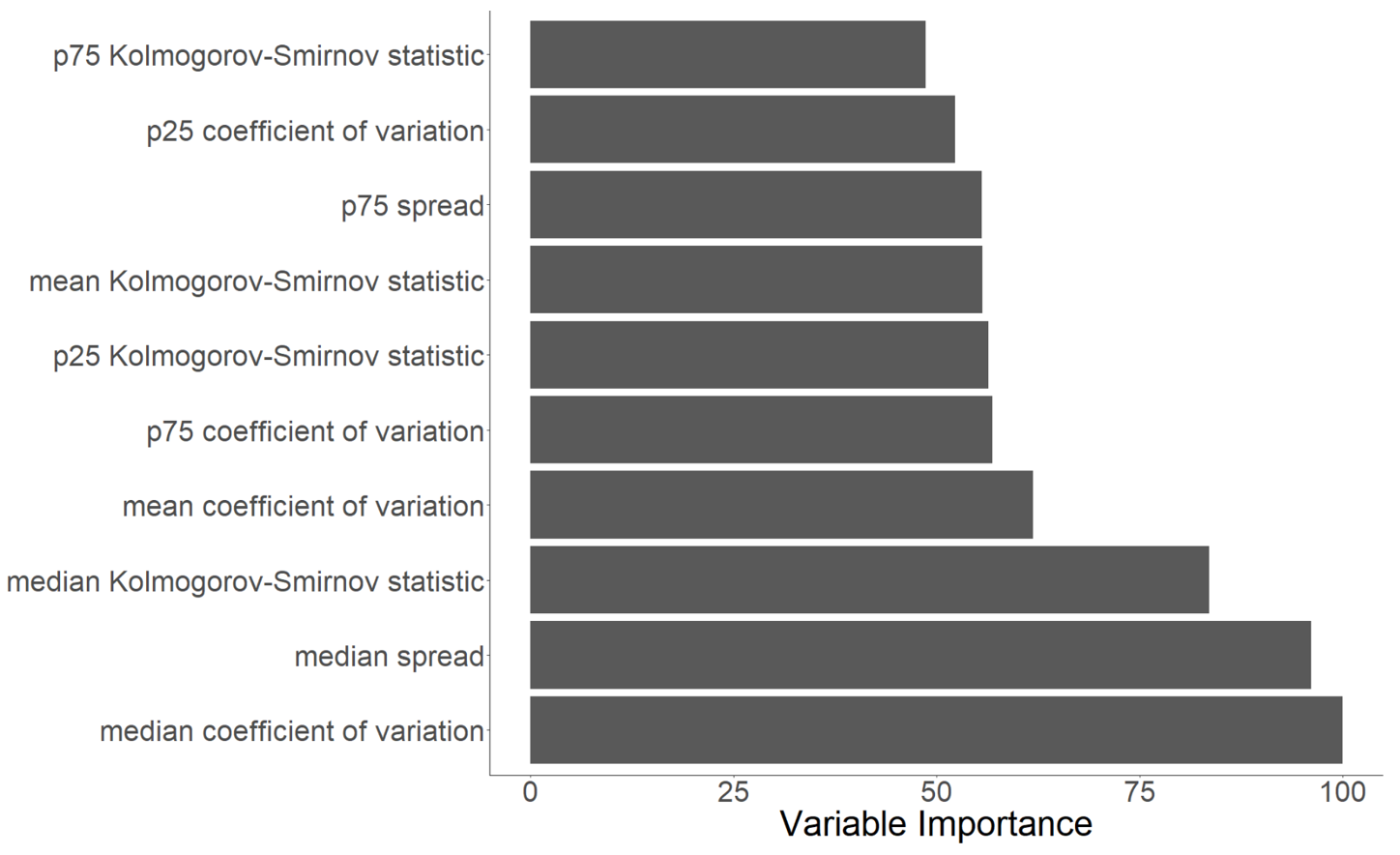}
	\end{figure}
	
	By looking at the three most important predictors according to the random forest, we find the medians for the coefficient of variation, the spread and the KS statistic remain the most predictive coalition-based screens (see Figure \ref{VarImp_MorePred}). The upper and lower quartiles of these descriptive statistics appear in the top-ten best predictors, but rather not at the top of the ranking.

	\subsection{The investigation of predictors measuring asymmetry}
	
	In the three different countries, predictors measuring asymmetry do not appear to be important (according to the random forest) in flagging collusive coalitions when also implementing coalition-based screens related to the variance and the uniformity of bids. This result might be puzzling when we remember that \cite{imhof2019detecting}, \cite{huber2020transnational} and \cite{wallimann2020machine} find screens for the asymmetry to be relevant in predicting the Japanese and Swiss bid-rigging cartels. In fact, asymmetry in the distribution of bids arises when we simultaneously analyze the bids from the winner designated by the cartel and the cover bids submitted by other the cartel participants. In our coalition-based approach, however, we select only three bidders and thus not necessarily the designated winner. Therefore, the absence of the designated winner in calculating the tender-based screens with only three cartel participants can limit the predictive power of screens based on the asymmetry of bids.
	
	\begin{table}[ht]
		\caption{Changes in accuracy when only considering screens for the asymmetry as predictors}\label{CCRonlyasym}
		\begin{center}
			\begin{tabular}{lccc}
				\hline
				\multirow{2}{*}{Classifier} & \multicolumn{3}{c}{Changes in percentage points}               \\ \cline{2-4} 
				& CCR  & CCR collusion  & CCR competition \\ \hline
				Lasso                       & -2.9    & -2.2              & -1.3                \\
				Random   forest             & -2.7    & -1.4              & -4.1                \\
				Super learner               & -2.3    & -1.8              & -2.7               \\
				Support   vector machines   & -3.2    & -2.3              & -3.4               \\ \hline
			\end{tabular}
		\end{center}
		\par
		\textit{Note: 'CCR' denotes the correct classification rate, 'CCR collusion' the correct classification rate of the collusive coalitions, and 'CCR competition' the correct classification rate of competitive coalitions.}
	\end{table}
	
\begin{figure}[!htp]
		\centering \caption{\label{VariImp_Onlyasym} Variable importance plot for the Swiss cartels with only screens for the asymmetry of bids. We compute the variable importance using the mean decrease in the Gini index, and express it relative to the maximum.}
		\includegraphics[scale=0.45]{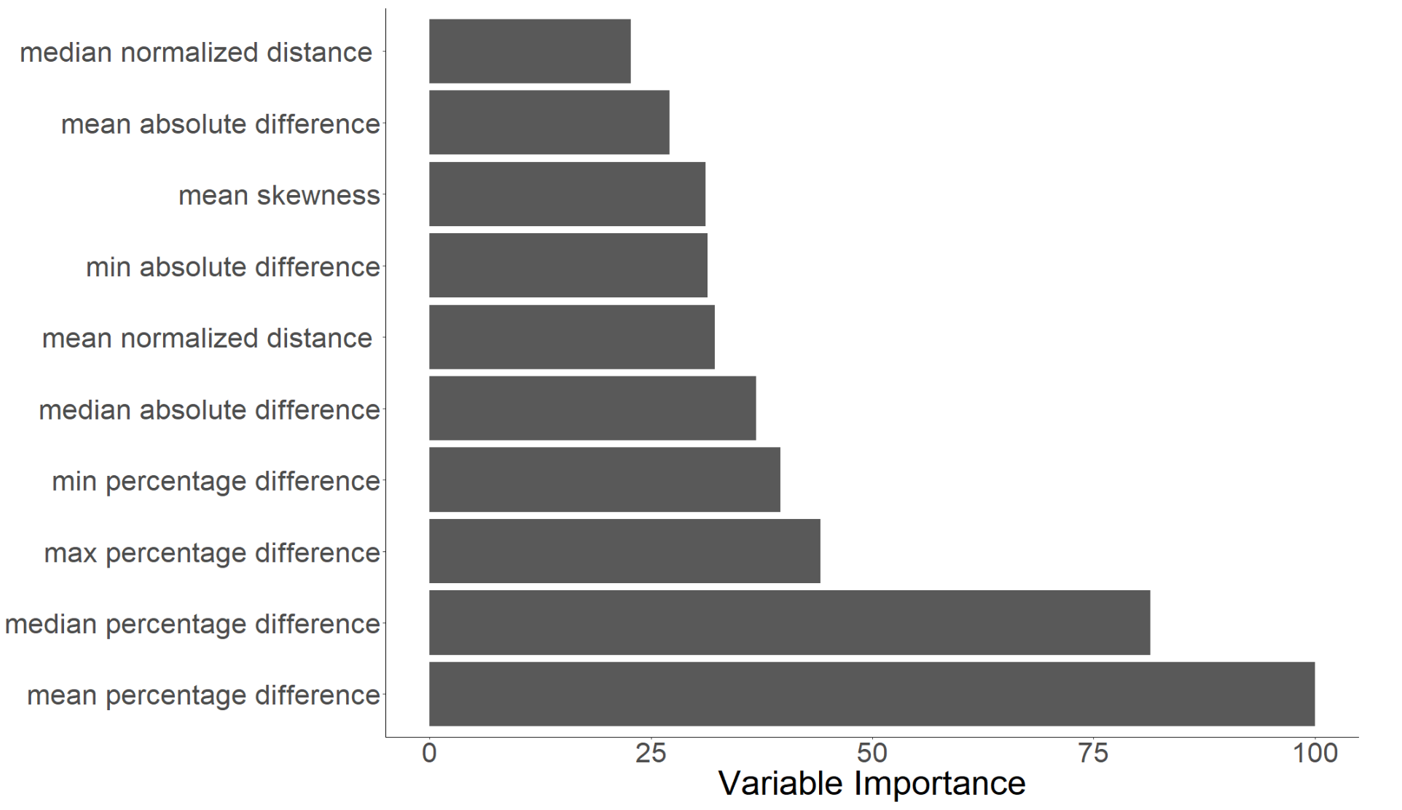}
	\end{figure}

	To investigate the importance of these screens further, in a second complementary analysis we discard screens for the variance and the uniformity of bids. We then repeat our estimation procedure for the Swiss data. Table \ref{CCRonlyasym} reports that correct classification rates decrease by 2.3 to 3.2 percentage points. Figure \ref{VariImp_Onlyasym} shows that summary statistics for the percentage difference are the most predictive coalition-based screens. However, these coalition-based screens are related to the variance of the bids if they do not include the designated winner’s bid. If variance is reduced for the losing bids, and if one takes into account mainly the losing bids in calculating the tender-based screens, then the coalition-based screens for the percentage difference will be smaller for collusive coalitions than for competitive coalitions. In fact, a look at the descriptive statistics indicates that the mean of the Swiss cartels' medians of the percentage difference for the collusive coalitions amounts to 3.3, as opposed to 5.6 for the competitive coalitions.
	
	Therefore, in a second step, we discard coalition-based screens for the percentage difference and the absolute difference since they might be related to the variance screens in order to analyze only screens for the asymmetry of the bids. Using sixteen predictors for the asymmetry of bids, we obtain a considerable decrease in correct classification rates of from 17.3 to 20.5 percentage points (see Table \ref{CCRonlyasym_discPercAbsol}). The decline is less for collusive coalitions (from 11.8 to 20.0 percentage points) but still large. In conclusion, coalition-based screens for asymmetry do not seem to be important for flagging collusive coalitions. The variance and the uniformity of bids therefore remain the most important features for describing changes in the distributional pattern of the bids in collusive coalitions.

	\begin{table}[ht]
		\caption{Changes in accuracy when only considering screens for the asymmetry as predictors and discarding screens for the percentage difference and the absolute difference }\label{CCRonlyasym_discPercAbsol}
		\begin{center}
			\begin{tabular}{lccc}
				\hline
				\multirow{2}{*}{Classifier} & \multicolumn{3}{c}{Changes in percentage points}               \\ \cline{2-4} 
				& CCR  & CCR collusion  & CCR competition \\ \hline
				Lasso                       & -17.7   & -15.2             & -18.7                \\
				Random   forest             & -21.2   & -19.8             & -23.9               \\
				Super learner               & -20.5   & -20.0             & -23.0              \\
				Support   vector machines   & -18.0   & -11.8             & -24.2              \\ \hline
			\end{tabular}
		\end{center}
		\par
		\textit{Note: 'CCR' denotes the correct classification rate, 'CCR collusion' the correct classification rate of the collusive coalitions, and 'CCR competition' the correct classification rate of competitive coalitions.}
	\end{table}

	\subsection{Coalition-based screens with four bidders}
	
	In a last step, we investigate the correct classification rate by forming coalitions of four firms, not three. However, it is more advisable to compute coalitions of three firms since it allows bid-rigging cartels formed with three bidders to be uncovered. Using coalitions with four firms makes it difficult to detect bid-rigging cartels formed with three bidders and would therefore restrict the broader scope of application of our suggested method based on coalitions. Moreover, calculating coalitions based on four firms might be somewhat more intense computationally. For example, if we calculate all possible coalitions of three firms formed with 75 firms, we obtain 67,525 potential coalitions to calculate. With coalitions of four firms for 75 firms, the potential coalitions amount to 1,215,450, or eighteen times the number of potential coalitions with three firms.
	
	Nonetheless, we calculate coalitions of four firms using the Swiss data, which as a computation perspective is easier, since there are three datasets for each cartel with a lower number of firms than the other cartels in Italy and in Okinawa. We end up with a total of 3,207 coalitions of four firms, 2,097 collusive and 1,110 competitive coalitions. We then use the same coalition-based screens as in Section \ref{empanaly} to recapitulate the changes in percentage points for the correct classification rates (correct classification rates for coalitions of four firms minus correct classification rates for coalitions of three firms from Table \ref{CCRSwisscart}). We find that the overall correct classification rates for coalitions of four firms is a little higher, with an increase of from 2.1 to 4.0 percentage points (see Table \ref{CCR_Four}), compared to the correct classification rates for three coalitions. It seems that the increase is mainly driven by an increase in the correct classification rates of collusive coalitions amounting to from 5.1 to 6.1 percentage points. The correct classification rates for competitive coalitions at the opposite fall by 1.8 to 2.3 percentage points except for the super learner, with a small increase of 0.9 percentage points. We also observe that coalition-based screens for asymmetry in the form of the skewness of the bids appear in the top-ten predictors according to the random forest with the coefficient of variation. Including a higher number of firms in the coalition reduces the likelihood that the winning bids will be omitted and therefore includes a greater distance between the first and second lowest bids in the coalition, naturally leading to more asymmetry. The fact that coalition-based screens for the asymmetry of bids have a higher predictive power could explain the overall rise in the correct prediction rates, specifically those of the collusive coalitions. The increase in the correct classification rates for coalitions with four firms also appears unsurprising since in most tenders in the Swiss data there were four or more cartel participants. 
	
	\begin{table}[ht]
		\caption{Changes in accuracy when using coalition-based screens with four bidders in the Swiss data}\label{CCR_Four}
		\begin{center}
			\begin{tabular}{lccc}
				\hline
				\multirow{2}{*}{Classifier} & \multicolumn{3}{c}{Changes in percentage points}               \\ \cline{2-4} 
				& CCR  & CCR collusion  & CCR competition \\ \hline
				Lasso                       & 3.4   & 5.4             & -1.8                \\
				Random   forest             & 2.1   & 5.1             & -2.3               \\
				Super learner               & 4.0   & 5.9             & 0.9              \\
				Support   vector machines   & 3.6   & 6.1             & -1.9              \\ \hline
			\end{tabular}
		\end{center}
		\par
		\textit{Note: 'CCR' denotes the correct classification rate, 'CCR collusion' the correct classification rate of the collusive coalitions, and 'CCR competition' the correct classification rate of competitive coalitions.}
	\end{table}

	\section{Policy recommendations}\label{policyrecom}
	
	\subsection{Advantages of a coalition-based detection method}
	The coalition-based approach proposed in this paper has several advantages in flagging bid-rigging cartels. We first reach correct classification rates of 90\% with the super learner in Italy, Japan and Switzerland. In other words, we classify nine coalitions out of ten correctly on average. This result remains stable, also while considering different auction formats, i.e., the first-price sealed-bid and the average bid auction. Super learner outperforms the other algorithms in two out of three cases and does not exhibit an imbalance in predicting both classes (collusive and competitive coalitions). Its greater performance derives from the use of multiple machine learning models, for which the algorithm creates an optimal weighted average.\footnote{See also https://cran.r-project.org/web/packages/SuperLearner/vignettes/Guide-to-SuperLearner.html (accessed 30 April 2021).}  Super learner is then advisable in our case. Besides, the machine learning literature is rapidly growing, and we assume that future research implementing novel machine learning algorithms will increase accuracy.
	
	Moreover, our coalition-based approach directly flags firms as cartel participants and is able to detect complete and incomplete bid-rigging cartels. If correctly calibrated, it can also flag partial cartels, that is, complete or incomplete bid-rigging cartels active in one specific area or one specific type of contract \citep[see, for example,][]{imhof2018screening,Abrantes2006}. In our cases, the bid-rigging cartels in Japan and in Switzerland are complete for a majority of tenders, but the Italian bid-rigging cartels are not. Identifying sub-groups of firms as cartel participants is important because markets are not always characterized by bid-rigging conspiracies affecting all contracts or involving all the firms. Therefore, our approach is not only applicable to different countries or auction formats, but also to different kind of bid-rigging cartels. Such possible broad applications render the coalition-based approach attractive for screening procurement markets and future research.
	
	Finally, the data requirement is low, as we need only the bids and the identity of the firms to calculate coalition-based screens. Other tender-based screens, such as those dealt with in \cite{huber2019machine} or \cite{wallimann2020machine}, do not require the bidders to be identified. Such low data requirements contrast with other methods of detection, which need cost-related variables or firm-specific covariates to implement the econometric tests, as suggested in \cite{Bajari2003} or more recently in \cite{conley2016detecting}. A low data requirement is crucial for two reasons. First, it allows the screening of large procurement datasets. If the data are available in a digital form, a competition or procurement agency can apply the detection method in a minimum amount of time. Second, it could be difficult to obtain information specific to firms without attracting the cartel’s attention to a possible investigation. Indeed, in some cases it could destabilize the cartel and have a preventive effect. However, cartel participants will certainly take more precautions and destroy evidence impeding the success of a future investigation.
	
	\subsection{Ex-ante Screening}
	
	When screening procurement markets, we suggest two different possibilities. The first consists of using data from previous cartels to fit predictive models (with machine learning algorithms) in order to apply them to a new dataset for which no prior information on collusion exists. The second possibility is to use benchmarks to isolate groups of suspicious contracts or firms. For the latter possibility, Table \ref{Coalitionsmeans} in Section \ref{empanaly} might offer a starting point for screening procurement markets. 
	
	However, for both possibilities, one should be aware that the institutional context of each country – for example the choice of the auction format or other country-specific characteristics – largely influences the distribution of bids in each tender. Coalition-based screens thus exhibit dissimilar values across countries and classes. For example, the values of the coefficients of variation for the Swiss collusive coalition exhibit slightly higher average values than the Japanese competitive coalitions (see Table \ref{Coalitionsmeans}). Therefore, training models in one country to be able to test them in another could in such circumstances be hazardous, as already noted by \cite{huber2020transnational}. Nonetheless, the effects of bid rigging go in the same direction, and their magnitudes might be similar in some cases. Hence, if a competition agency intends to apply the method to a different market or country, we recommend using benchmarks based on the effect of bid rigging rather than benchmarks based on the level of the screens. For example, a decrease by two in the variance on a market could be suspicious and should be subjected to further statistical inquiry to confirm the initial diagnostic. Moreover, further research should investigate the possibility of normalizing bids or screens by country to enable predictive models to be transferred directly from one country to another.
	
	A competition agency can implement both predictive tender-based and coalition-based screens to fit models or assess approximate benchmarks. If the amount of data to screen is small (e.g., fewer than hundred firms bidding in the data), one can directly apply the coalition-based approach. However, if the amount of data to screen is large (e.g., more than a thousand firms bidding in the data), the tender-based approach of \cite{huber2019machine} or \cite{wallimann2020machine} is simpler to apply (e.g., less computationally intensive) in order to identify markets for specific products or different geographical areas that are potentially suspicious.
	
	To increase the confidence level, a competition agency could also combine both types of screens, i.e., tender-based and coalition-based. Once a bench of suspicious tenders with the tenders-based screens has been identified, one can apply the coalition-based screens to verify whether the firms participating in the suspicious tenders are sufficiently suspect to open an investigation. Such a double testing procedure increases the reasonable grounds for identifying bid-rigging conspiracies and offers greater confidence to competition agencies in screening procurement markets. Here the coalition-based approach will provide precious assistance because it allows the identity of potential cartel participants to be confirmed with a high degree of confidence: the correct prediction rates in the three different countries indicate that nine firms out of ten are correctly classified as being competitors or cartel participants. In other words, a firm flagged as potentially collusive using the coalition-based approach has a 90\% likelihood of being a cartel participant. The level of likelihood should be sufficiently high to constitute reasonable grounds for opening an investigation.

	\section{Conclusion}\label{Conclusion}
	Our paper contributes to the literature on cartel detection in manifold ways. We have developed an original detection method based on screens by focusing on coalitions. This approach allows a broader application by detecting complete and incomplete bid-rigging cartels as well as partial cartels in different auction formats. Coalition-based screens delivered more correct classification rates than previous methods using tender-based screens and, using the super learner, correctly classified on average at least nine coalitions out of ten in Italy, Japan and Switzerland. The performance of the super learner surpassed other algorithms and is balanced across collusive and competitive coalitions. It thus remained the most suitable algorithm for our application.
	
	Although an increase in the performance of three to ten percentage points might appear low, the coalition-based screens reduced the error rate by half in some cases. Such falls in the error rate are desirable given the heavy legal and procedural consequences for firms that have been flagged as potential cartel participants. Furthermore, the coalition-based screens do not oppose the tender-based screens but constitute an interesting complement limiting the risk of false positives in screening procurement markets. 
	
	Furthermore, we found that the levels of the most important coalition-based screens differ considerably between countries, though the magnitude of the effects of bid rigging bids is similar. Thus, a decrease by a factor of two in the median of the coefficient of variation and the spread, as well as an increase by a factor of two in the median of the KS statistic, could indicate potential collusion. Future empirical research should investigate the possibility of normalizing screens or bids by country or market to continue developing a general screening method that is both the most reliable and has the broadest applicability. In addition, future theoretical research should focus on structural models explaining why bid rigging reduces the variance and renders the distribution of bids less uniform than in competitive tenders.

	\newpage
	\bigskip
	
	\bibliographystyle{econometrica}
	\bibliography{coalition}
	
	\bigskip
	
\end{document}